\documentclass[aps,prc,showpacs,showkeys,superscriptaddress]{revtex4}
\usepackage{amsfonts}
\usepackage{amsmath}
\usepackage{amssymb}
\usepackage{graphicx}
\usepackage{subfigure}

\begin{document}

\title{Enhancement of sub-barrier fusion of two-neutron halo nuclei }
\author{P.\ R.\ S.\ Gomes}
\email{paulogom@if.uff.br}
\affiliation{Instituto de F\'{\i}sica, Universidade Federal Fluminense, Av. Litoranea,
24210-340, 21941-972 Niteroi, Brazil }
\author{L.\ F.\ Canto}
\email{canto@if.ufrj.br}
\affiliation{Instituto de F\'{\i}sica, Universidade Federal do Rio de Janeiro, C.P.\
68528, 21941-972 Rio de Janeiro, RJ, Brazil}
\author{J. Lubian}
\email{lubian@if.uff.br}
\affiliation{Instituto de F\'{\i}sica, Universidade Federal Fluminense, Av. Litoranea,
24210-340, 21941-972 Niteroi, Brazil }
\author{M.\ S.\ Hussein}
\email{hussein@if.usp.br}
\affiliation{Instituto de F\'{\i}sica, Universidade de S\~{a}o Paulo, C.P.\ 66318,
05314-970 S\~{a}o Paulo, SP, Brazil.}

\begin{abstract}
The tunneling of composite systems, where breakup may occur during the
barrier penetration process is considered in connection with the fusion of
halo-like radioactive, neutron- and proton-rich nuclei on heavy targets. The
large amount of recent and new data clearly indicates that breakup hinders
the fusion at near and below the Coulomb barrier energies. However, clear
evidence for the halo enhancements, seems to over ride the breakup hindrance
at lower energies, owing, to a large extent, to the extended matter density
distribution. In particular we report here that at sub-barrier energies the
fusion cross section of the Borromean two-neutron halo nucleus $^{6}$He with
the actinide nucleus $^{238}$U is significantly enhanced compared to the
fusion of a no-halo $^{6}$He. This conclusion differs from that of the
original work, where it was claimed that no such enhancement ensues. This
sub-barrier fusion enhancement was also observed in the $^{6}$He + $^{209}$%
Bi system. The role of the corresponding easily excitable low lying dipole
pygmy resonance in these systems is therefore significant. The consequence
of this overall enhanced fusion of halo nuclei at sub-barrier energies, on
stellar evolution and nucleosynthesis is evident.
\end{abstract}

\maketitle

Fusion processes between heavy ions have been a subject of major interest in
the last four decades. Two major motivations are the potential production of
super heavy elements which do not exist in nature, and the understanding of
stellar evolution and nucleosynthesis. At energies below the Coulomb
barrier, tunneling becomes the only means for fusion to occur. This purely
quantum mechanical effect predicted more than 70 years ago by Gamow, remains
a subject of great interest for theorists. What happens when the tunneling
systems are composite? How the internal structure of system modifies Gamow's
theory? These are important questions for which nuclear physics can supply
clear answers. In recent years the fusion of extended nuclear systems such
as neutron and proton rich halo isotopes has attracted great interest \cite%
{CGD06,Hu91,TS91,HPC92,BFZ07,AH09}. Two competing effects seem to operate in
such cases. The dynamic effects related to the very low threshold for
breakup of the halo nucleons, and the static effects of the extended matter
distribution of these same nucleons. The major question is how to clearly
identify these effects and check how they individually influence the
tunneling/fusion at low energies. One guiding principle used to answer this
question is that the dynamic breakup coupling effect is strongly energy
dependent and dispersive while the static effects are mostly accounted for
through the use of the proper matter density distribution in the
construction of the overall double folding interaction potential, an
energy-independent entity.

Here we present clear evidence of the enhancement of the fusion probability
of the Borromean nucleus $^{6}$He with the heavy targets $^{209}$Bi and $%
^{238}$U at sub-barrier energies. This is in contrast to the conclusions
reached by Raabe et al. \cite{RSC04}, where it was claimed that no such
enhancement ensues in the $^{6}$He + $^{238}$U system. As we show below, a
proper account of the static effects of the halo as manifested in the use of
the correct matter density of $^{6}$He used in the double folding model and
used to calculated the tunneling probability with the coupled channels model
containing the dynamic breakup coupling effects would lead to a fusion cross
section which contains the above mentioned enhancement. What Raabe et al.
did was to compare their data with such theory, which would only show
concurrence and would lead to the conclusion they reached. One would need to
make a comparison with a theory which does not contain the halo static
effects in order to assess the matter. In the following we do exactly this.

A large body of fusion data is available which can be used to define
unambiguously what we call the "background" with which a sensible comparison
can be made in order to decide whether the halo fusion system exhibits
enhancement at sub-barrier energies or not. How to present this background
tunneling/fusion data of the many complex many-body systems alluded to above
is a subtle question which has recently been addressed and with success \cite%
{CGL09,CGL09b}. The idea is to define a universal fusion function (UFF),
which accounts for most of the, properly scaled, fusion data. This UFF is
defined by first transforming the collision energy and the fusion cross
section into dimensionless quantities, according to the prescription

\begin{equation}
F_{exp}(x)=\frac{2E}{\hbar \omega R_{B}^{2}}\ \sigma _{F}  \label{bench1}
\end{equation}%
where, $x=\frac{E-V_{B}}{\hbar \omega }$, with, $R_{B}$, $V_{B}$ and $\hbar
\omega $, being the barrier radius, height, and curvature parameters of the
fusion (Coulomb) barrier, respectively. One can determine the optical fusion
function, $F_{opt}(x)$, using the cross section $\sigma _{F,opt}$ predicted by
the optical model calculation with a potential which leads to those barrier
parameters. This fusion function is system-independent when $\sigma _{F,opt}$
is accurately described by Wong's formula \cite{Won73}. In this case 
\begin{equation}
F_{opt}(x)\rightarrow F_{0}(x)=\ln \left[ 1+\exp \left( 2\pi x\right) \right]
\label{bench2}
\end{equation}

The function $F_{0}(x)$ is universal as it is independent on the fusing
system. At very small values of $x$, $F_{0}(x)\rightarrow e^{2\pi x}$, while
at large $x$, it acquires the simple linear form, $F_{0}(x)\rightarrow 2\pi
x $. It has the conspicuous limiting value of $\ln 2$ at $x=0$, namely at a
center of mass energy equal the Coulomb barrier height. These
characteristics makes $F_{0}(x)$ a quite convenient benchmark to which
reduced data are compared.

The first step to use this method in the analysis of the fusion data is to
build the experimental fusion function, $F_{exp}(x)$. This is done using the
experimental fusion cross section defined before. However, as in most cases
the fusion cross section is strongly affected by channel couplings and
Wong's model is not exact for light systems and at sub-barrier energies, one
introduces \cite{CGL09,CGL09b} a renormalized experimental fusion function, 
\begin{equation}
{\bar{F}}_{exp}(x)=\frac{F_{exp}(x)}{\left[ \frac{\sigma _{F,cc}^{A}}{\sigma
_{F,opt}}\right] }.  \label{UFF}
\end{equation}

Here, $\sigma _{F,opt}$ is the theoretical fusion cross section with all
couplings switched off, and $\sigma _{F,cc}^{A}$ is the cross section
obtained from a coupled channel ($CC$) calculation including a set of
channels $A$. If all relevant channels are included in $A$ and the correct
coupling strengths are used, the renormalized experimental fusion function,
Eq.(\ref{UFF}), should match the benchmark, Eq.(\ref{bench2}), namely, 
\begin{equation}
{\bar{F}}_{exp}(x)\rightarrow F_{0}(x)
\end{equation}

If on the other hand some relevant set of channels $B$ is left out of the $CC
$ calculation then ${\bar{F}}_{exp}/F_{0}=\sigma _{F,cc}^{A+B}/\sigma
_{F,cc}^{A}$. Clearly, the ratio ${\bar{F}}_{exp}/F_{0}$, would give a
precise measure of the importance of the left out channels not included in
the $CC$ calculation $A$. Through this procedure, one is able to isolate the
effect of the breakup channel coupling on the fusion cross section of weakly
bound systems using as a theoretical model a $CC$ calculation involving
bound channels only ($A$). However, the inclusion of transfer channels in
calculations of $\sigma _{F,cc}$ may be a difficult task, specially in
collisions with large positive transfer $Q$-values, such as neutron transfer
in collisions with $^{6}$He. When transfer channels are important and they
are not included in the $CC$ calculation, the differences between ${\bar{F}}%
_{exp}(x)$ and the $F_{0}(x)$ should be assigned to the combined effects of
couplings to breakup and transfer channels.\newline

We now use this method of analysis to take a new look at "old" data.\newline

In order to make a sensible description of the fusion of weakly bound
nuclei, it has been customary to distinguish between the complete fusion
(CF) where the whole projectile is captured by the target, and the
incomplete fusion (ICF), where a fragment of the broken projectile is
captured. The total fusion (TF) is then defined as the sum of these two
cross sections \cite{CGD06}. In many instances it has been rather difficult
to separate experimentally these two components. The same can be said about
the theoretical view point. The widely used Continuum Discretized Coupled
Channels (CDCC) model can calculate the total fusion cross section, the
total absorption cross section form the continuum channels, but it fails to
supply an unambiguous way to calculate the ICF of a given fragment \cite%
{DiT02}. \newline

In Fig. 1, we show the renormalized experimental fusion function ${\bar{F}}%
_{exp}(x)$, of Eq.(\ref{UFF}), vs. $x$, for several tightly and weakly bound
systems. The results are shown in a logarithmic scale. We see clearly that
the "data" for the tightly bound systems follow quite closely the Universal
Fusion Function (UFF), $F_{0}$, indicating that the chosen $A$ channels are
adequate to describe the fusion of these systems, as has been emphasized in 
\cite{LDH95,MBD99}. The results for the weakly bound systems indicate strong
deviations from the UFF at above-barrier energies. This deviation becomes
quite visible if the data are presented on a linear scale. This is shown in
Fig. 2. For the $TF$ for one stable weakly bound system ($^{9}$Be +$^{208}$%
Pb) \cite{DHB99,DGH04} and for another, unstable, bound system ($^{17}$F + $%
^{208}$Pb) \cite{REJ98}, there is good agreement above the barrier ($x$ $>$
0) and a slight enhancement at sub-barrier energies ($x<$ 0). The CF for the
same $^{9}$Be + $^{208}$Pb system and for the $^{6,7}$Li + $^{209}$Bi
systems \cite{DGH04}\ is suppressed by about 30$\%$ at energies above the
barrier. Thus, this suppression is attributed to the loss of flux going to $%
ICF$, following breakup. The fusion enhancement at sub-barrier energies is
attributed to prompt and resonance breakups and transfer channels. For
fusion induced by $^{6}$He \cite{RSC04, KGP98}, there is also a significant
suppression above the barrier and some slight enhancement at sub-barrier
energies. However, for this latter very weakly bound halo nucleus induced
fusion system, only $TF$ was measured. As sharing energy considerations \cite%
{CGL09} show that the $ICF$ of $^{4}$He with the targets is unlikely, the
fusion suppression for the neutron halo systems is attributed to transfer
and/or non-capture breakup channels, rather than to $ICF$. \newline

We now turn to the important question of how does the coupling to the
breakup channel competes with the extended matter distribution, which is
expected to play a major role, as emphasized by several authors? Even with
extensive experimental and theoretical efforts during the last two decades 
\cite{CGD06, Hu91, TS91, HPC92, BFZ07, AH09}, a full understanding of the
competing effect of the breakup coupling and the halo is still not fully
reached though great progress towards this goal have been made. The reason
behind this is the lack of a full exact theory of three- and four-body
nature of the reactions of one- and two- nucleon halo nuclei, and how
tunneling of a structured system is described. \newline

We first dwell on the dynamic effects of the halo, exemplified by the
coupling to the breakup channel. This is conveniently described through the
dynamic polarization potential (DPP) which represents the Feshbach reduction
of a $CC$ description into an one effective channel description. The
breakup, dispersive, energy-dependent DDP for weakly bound systems, $%
V_{bu,pol}(E)$, has been extensively studied within the CDCC, and the
conclusion reached, \cite{LuN07, CHC08, CLG09}, is that its imaginary part
was found to suffer a slight increase as the energy is lowered below the
barrier, followed by a drop to zero as the break up channel closes. The real
part of the $V_{bu,pol}$ was found to be repulsive (positive) in the barrier
region. The over all effect of the breakup DPP, $V_{bu,pol}$, is to induce a
reduction in the fusion at energies above the barrier due to a large extent
to the rather long range nature of "dynamic absorption" described by $%
ImV_{bu,pol}$, and to the repulsive $ReV_{bu,pol}$. The effect of $%
V_{bu,pol} $ below the barrier is rather small. In contrast, the static
effect of the halo, present in the bare optical potential described by an
appropriate double-folding model, is always present at all energies. In
figure 3 we show the barrier calculated with and without the halo, for the $%
^{6}$He + $^{238}$U\ system. Clearly the halo makes the barrier lower. The
overall effect is a larger penetrability when compared to the no halo
barrier.\newline

The connection between the imaginary and real parts of the DDP is dictated
by the dispersion relation. The results discussed above concerning the
breakup DPP, $V_{bu,pol}(E)$, has been referred to as the Breakup Threshold
Anomaly (BTA), \cite{HGL06}, quite useful in the analysis of elastic
scattering of weakly bound systems. In contrast, the DDP for bound channels,
which also obeys a dispersion relation, presents a real and imaginary parts
in the barrier region, with characteristics which are opposite to those of
the breakup DDP, namely, the real part is attractive while the imaginary
part drops as the energy is lowered below the barrier. This behavior has
come to be known as the Threshold Anomaly (TA).\newline

Guided by the above considerations we proceed now with a detailed discussion
of the fusion of $^{6}$He. In figure $4$ we show results of two coupled
channels calculations for the $^{6}$He + $^{238}$U system, toghether with
the data. We use as the bare potential the double --folding Sao Paulo
potential ($SPP$) \cite{CDH97,CCG02}. Data are from ref \cite{RSC04}. The
full curve is the result of the calculation which does not include the
effect of the two neutron halo of $^{6}$He. The folding potential here was
obtained by using the typical nuclear density of the strongly bound
projectile $^{4}$He, scaled to the size of $^{6}$He. In this case, the
differences with respect to the data originate from the absence, in the
calculation, of the static effect of the halo and couplings to the breakup
channel. When the static halo effects are included, by using the realistic $%
^{6}$He density in the construction of the $SPP$, the calculated fusion
cross section is increased, as seen in the dashed curve in Fig. 4. This
result agree with Raabe et al \cite{RSC04} and also show suppression of
fusion cross section above the barrier. As already emphasized, the dynamic
effects associated with the breakup channel, not included in either
calculations, result in a reduction of the fusion at above barrier energies,
and are insignificant at sub-barrier energies. Thus the static effect of the
halo is very important and enhances the fusion cross section in the whole
energy range, especially at sub-barrier energies. The above behaviour of the
fusion of Borromean nuclei such as $^{6}$He is independent on the target, as
it is shown in figure 5 for the $^{6}$He + $^{209}$Bi system \cite{KGP98}.
While we find similar static halo enhancement in the fusion of the two above
mentioned systems, the authors of the original papers reach opposing
conclusions. We believe that we have resolved this issue and put the problem
of the fusion enhancement of halo nuclei to rest.

In conclusion, we have presented convincing arguments to support our thesis
that the fusion of halo nuclei at sub-barrier energies is enhanced when
compared to the fusion of non-halo nuclei. This, together with the existence
of threshold dipole strength (the so-called pygmy resonances) would inflict
important changes in the scenario of the r-process in nucleosynthesis as
theoretically predicted by Goriely \cite{Go98}.

\bigskip \noindent \textbf{Aknowledgements}

\medskip \noindent This work was supported in part by the FAPERJ, CNPq,
FAPESP, PRONEX, and the Instituto Nacional de Ciencia e Tecnologia de
Informac\~{a}o Qu\'{a}ntica-MCT. We thank our collaborators, L. C. Chamon,
E. Crema, for fruitful discussions.\newline

\bigskip

\begin{figure}[tbp]
\begin{center}
\includegraphics*[height=10cm]{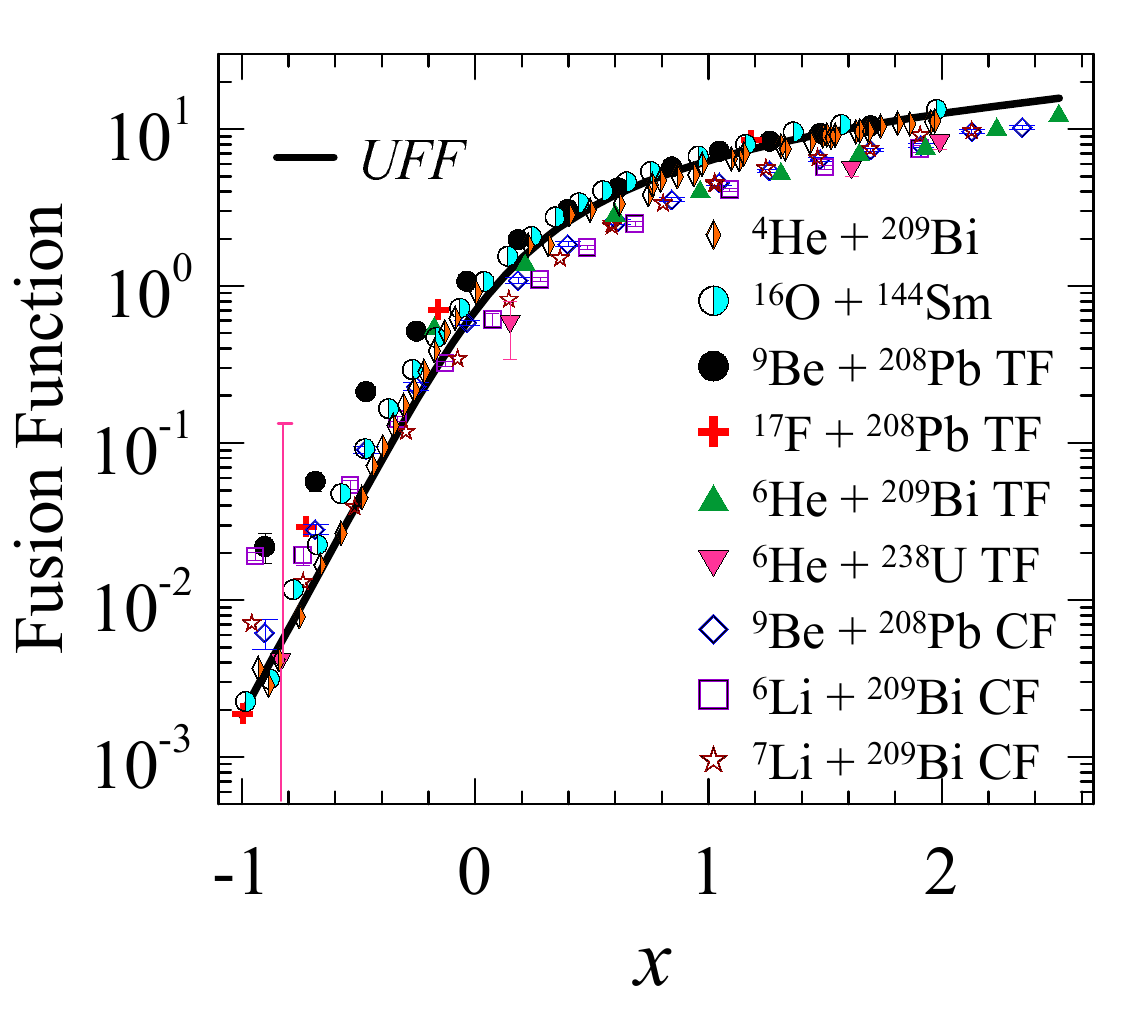}
\end{center}
\caption{The renormalized experimental fusion function ${\bar{F}}_{exp}(x)$,
of Eq.(\protect\ref{UFF}), vs. the variable $x$, for a sample of tightly
bound systems, $^{16}$O + $^{144,154}$Sm, and $^{4}$He + $^{209}$Bi, the
data points are, respectively from \protect\cite{LDH95,MBD99}, and for a
sample of weakly bound systems, $^{9}$Be + $^{208}$Pb, $^{17}$F + $^{208}$%
Pb, $^{6,7}$Li + $^{209}$Bi, $^{6}$He + $^{209}$Bi, and $^{6}$He + $^{238}$%
U. The full curve is the UFF, $F_{0}(x)$ of Eq. (\protect\ref{bench2}). The
data points are respectively from \protect\cite{LDH95} ($^{16}$O), 
\protect\cite{MBD99} ($^{4}$He), \protect\cite{DHB99,DGH04} ($^{9}$Be), 
\protect\cite{REJ98} ($^{17}$F), \protect\cite{DGH04} ($^{6,7}$Li), 
\protect\cite{KGP98, RSC04} ($^{6}$He)}
\label{fig1}
\end{figure}

\bigskip

\begin{figure}[ptb]
\begin{center}
\includegraphics*[height=10cm]{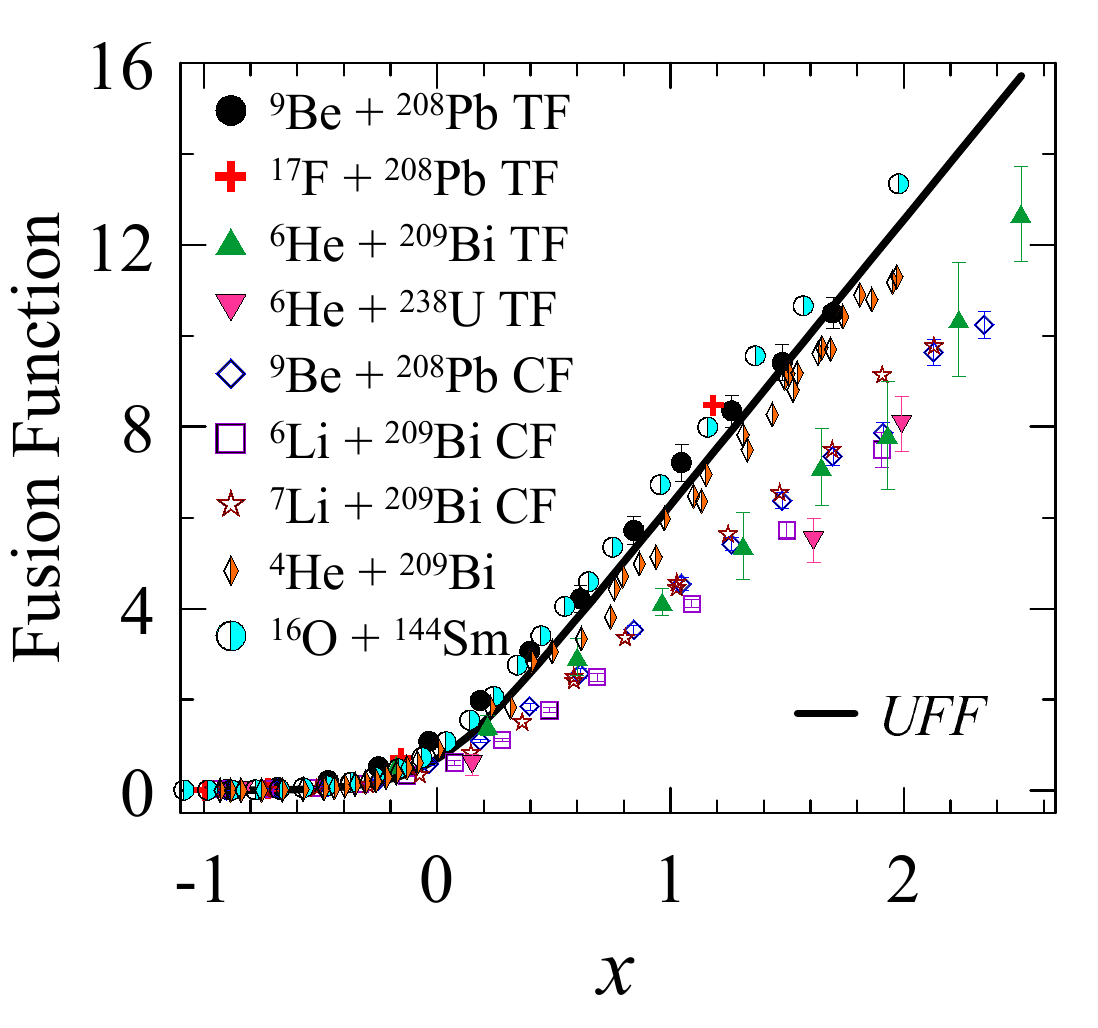}
\end{center}
\caption{Same as Fig. 1 on a linear scale}
\label{fig2}
\end{figure}

\bigskip

\begin{figure}[ptb]
\begin{center}
\includegraphics*[height=7cm]{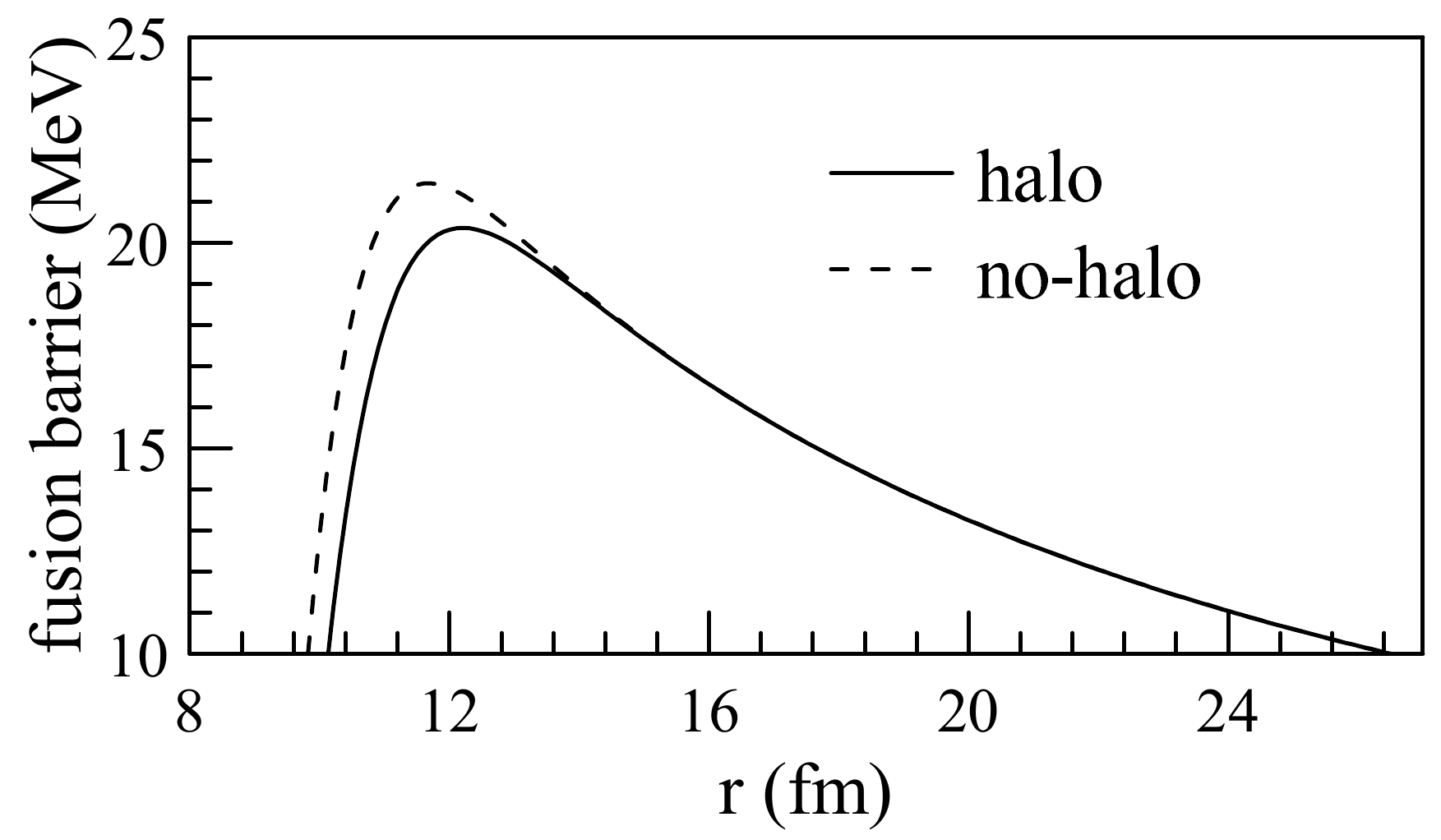}
\end{center}
\caption{The Coulomb barrier of the system $^{6}$He + $^{238}$U calculated
using the double folding-based S\~{a}o Paulo potential \protect\cite{CDH97,
CCG02}. The dashed curve is obtained with the actual realistic matter
density of $^{6}$He containing the two-neutron halo effect, while the full
curve shows the result obtained with a normal, non-halo, density of a $^{6}$%
He nucleus treated as an $\protect\alpha$ particle with six nucleons}
\label{fig3}
\end{figure}

\bigskip

\begin{figure}[ptb]
\begin{center}
\includegraphics*[height=10cm]{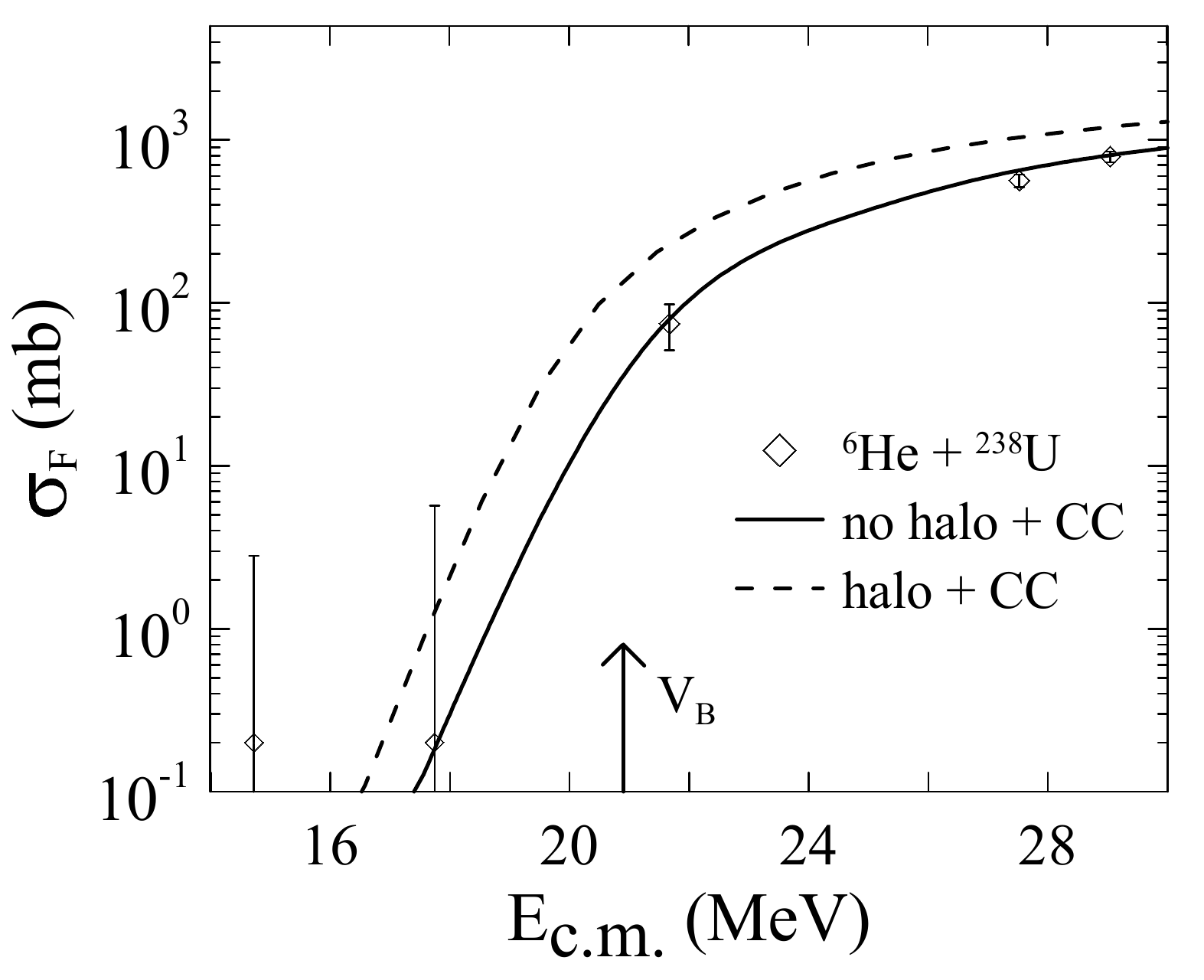}
\end{center}
\caption{shows results of two calculations of the fusion cross section for
the $^{6}$He + $^{238}$U system, together with the data of ref \protect\cite%
{RSC04}. The $SPP$ interaction of \protect\cite{CDH97,CCG02} is employed as
a background optical potential. The full curve is the result of the coupled
channels calculation which does not include the static effect of the two
neutron halo of $^{6}$He. The dashed curve is the corresponding coupled
channels result obtained with a background potential which contains the
static halo effect. The coupled channels included in both calculations
correspond to couplings to the main excitations of the target,$^{238}$U. }
\label{fig4}
\end{figure}

\bigskip

\begin{figure}[ptb]
\begin{center}
\includegraphics*[height=10cm]{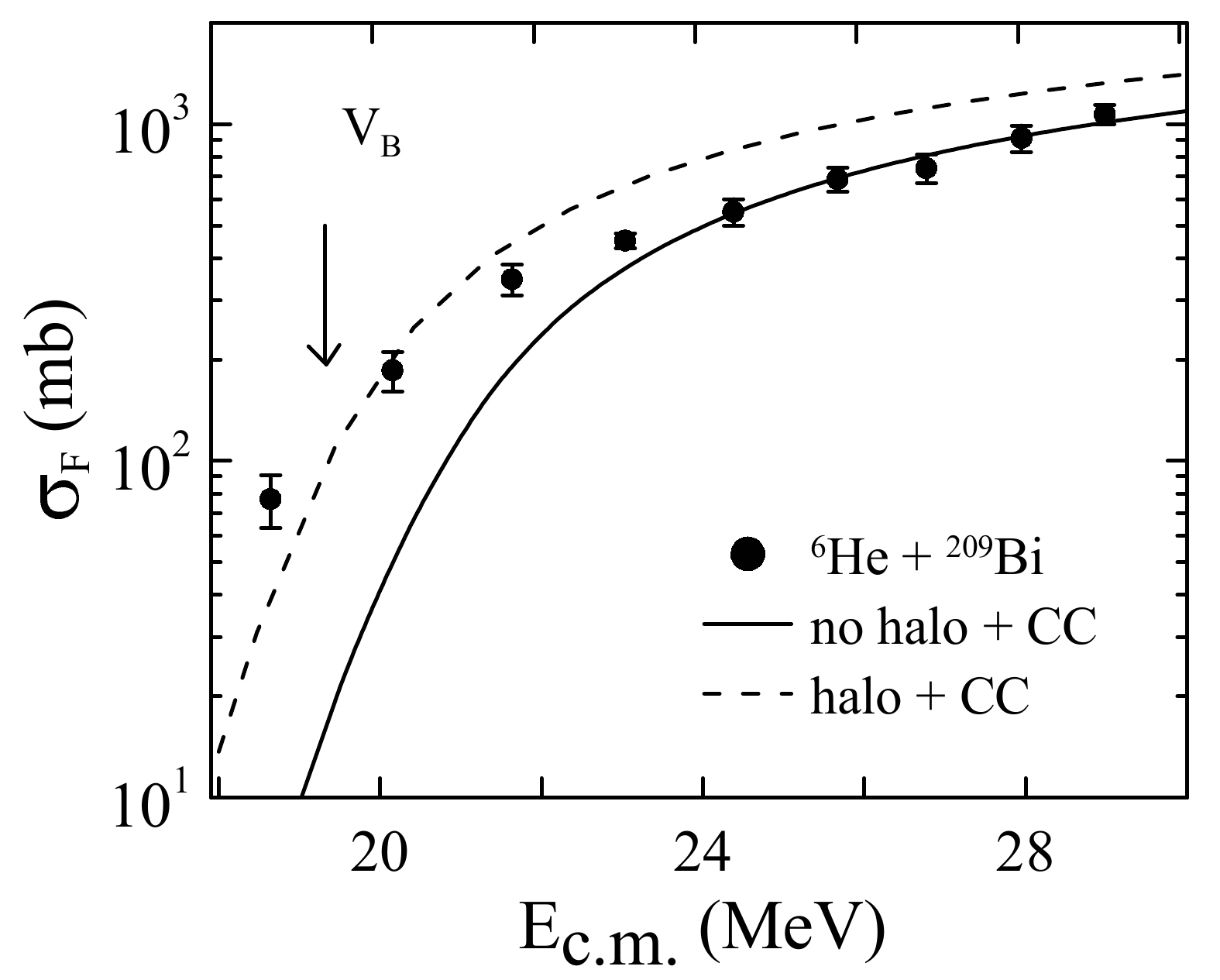}
\end{center}
\caption{Same as Fig. 4 for the system $^{6}$He +$^{ 209}$Bi. The data points are from ref. \cite{KGP98}. The theoretical curves corresponds 
to the ones of Fig. 4 as well.}
\label{fig5}
\end{figure}


\textbf{References}

\begin{thebibliography}{99}
\bibitem{CGD06} Canto, L. F., Gomes, P.R.S., Donangelo, R.\ \& Hussein, M.S., %
\newblock{\em Phys. Rep.}, 424 (2006) 1.

\bibitem{Hu91} Hussein, M.S. , \newblock{\em
Nuclear Physics A}, 531 (1991) 192.

\bibitem{TS91} Takigawa, N. \& Sagawa, H.,  \newblock{\em Phys. Lett. B}
265 (1991) 23.

\bibitem{HPC92} Hussein, M.S., Pato, M.P., Canto, L.F. \&Donangelo, R.,  %
\newblock{\em Phys. Rev. C}, 46 (1992) 377.

\bibitem{BFZ07} Bertulani, C.A., Flambaum, V.V. \& and Zelevinsky, V.G. J.,
Phys. G, 34 (2007) 2289.

\bibitem{AH09} Andrade, D.M. \&Hussein, M.S.,  \newblock{\em Phys. Rev. C},
80 (2009) 034610.

\bibitem{RSC04} Raabe, R. et al.,  \newblock {\em Nature}, 431 (2004) 823.

\bibitem{CGL09} Canto, L.F., Gomes, P.R.S., Lubian, J., Chamon, L.C. \&
Crema, E., \newblock{\em Nucl. Phys. A}, 821 (2009) 51.

\bibitem{CGL09b} Canto, L.F., Gomes, P.R.S., Lubian, J., Chamon, L.C. \&
Crema, E.,  \newblock{\em J. Phys. G}, 36 (2009) 015109.

\bibitem{Won73} Wong, C.Y.,  \newblock {\em Phys. Rev. Lett.}, 31 (1973) 766.

\bibitem{DiT02} Diaz-Torres, A. \&Thompson, I.J., \newblock {\em Phys. Rev.
C}, 65 (2002) 024606.

\bibitem{LDH95} Leigh, J.R. et al.,  \newblock {\em Phys. Rev. C},  52 (1995) 3151.

\bibitem{MBD99} Morton, C.R. et al.,  \newblock{\em Phys. Rev. C}, 60 (1999) 044608.

\bibitem{DHB99} Dasgupta, M. et al.,  \newblock {\em Phys. Rev. Lett.},
82  (1999) 1395.

\bibitem{DGH04} Dasgupta, M. et al.,  \newblock {\em Phys. Rev. C},
70 (2004) 024606.

\bibitem{REJ98} Rehm, K. E. et al.,  \newblock{\em
Phys. Rev. Lett.}, 81 (1998) 3341.

\bibitem{KGP98} Kolata, J.J. et al.,  \newblock {\em
Phys. Rev. Lett.}, 81 (1998) 4580.

\bibitem{CLG09} Canto, L.F., Lubian, J., Gomes, P.R.S. \& Hussein, M. S.,  
\newblock{\em Phys. Rev. C}, 80 (2009) 047601.

\bibitem{LuN07} Lubian, J. \& Nunes, F. M.,  \newblock{\em J. Phys. G}
34 (2007) 513.

\bibitem{CHC08} C\'{a}rdenas, W.H.Z., Hussein, M.S., Canto, L.F. \& Lubian,
J.,  \newblock{\em Phys. Rev. C} 77 (2008) 034609.

\bibitem{HGL06} Hussein, M.S., Gomes, P.R.S., Lubian, J. \& Chamon, L.C.,  
\newblock{\em Phys. Rev. C}, 73 (2006) 044610.

\bibitem{CDH97} Chamon, L.C., D. Pereira, D., Hussein, M.S., Ribeiro, M.A.C.
\& Galetti, D.,  \newblock{\em Phys. Rev. lett.} 79 (1997) 5218.

\bibitem{CCG02} Chamon, L.C. et al.,  \newblock {\em Phys. Rev. C},
66 (2002) 014610.

\bibitem{Go98} S. Goriely, S., \newblock{\em Phys. Lett. B}, 436 (1998) 10.
\end{thebibliography}
\end{document}